\documentclass[a4paper,useAMS,usenatbib,usegraphicx]{mn2e}
\usepackage{times}
\usepackage{xspace}
\usepackage{appendix}
\usepackage{amsfonts}
\usepackage{amsmath,amssymb,epsfig}
\usepackage{color}
\usepackage[dvipsnames]{xcolor}
\usepackage{alltt}
\usepackage[colorlinks=true,linkcolor=red,citecolor=blue,breaklinks=true]{hyperref}

\newcommand{\apjl}{ApJL}
\newcommand{\apj}{ApJ}

\newcommand{\mnras}{MNRAS}
\newcommand{\prd}{PRD}

\newcommand{\apjs}{ApJS}

\newcommand{\pasa}{PASA}
\newcommand{\aap}{AAP}

\newcommand{\jcap}{JCAP}

\newcommand{\healpix}{\ensuremath{\tt HEALPix}}
\newcommand{\synfast}{\ensuremath{\tt synfast}}
\newcommand{\healpy}{\ensuremath{\tt Healpy}}
\newcommand{\hammurabi}{\ensuremath{\tt Hammurabi}}
\newcommand{\polspice}{\ensuremath{\tt PolSpice}}

\newcommand{\Nside}{\ensuremath{N_{\rm side}}}
\newcommand{\Npix}{\ensuremath{N_{\rm pix}}}
\newcommand{\eor}{{\sc{EoR}}}

\def\deg{\ifmmode^\circ\else$^\circ$\fi}
\def\pdeg{\ifmmode $\setbox0=\hbox{$^{\circ}$}\rlap{\hskip.11\wd0 .}$^{\circ}
          \else \setbox0=\hbox{$^{\circ}$}\rlap{\hskip.11\wd0 .}$^{\circ}$\fi}
\def\arcs{\ifmmode {^{\scriptstyle\prime\prime}}
          \else $^{\scriptstyle\prime\prime}$\fi}
\def\arcm{\ifmmode {^{\scriptstyle\prime}}
          \else $^{\scriptstyle\prime}$\fi}

\newcommand{\hi}{H{\sc i}}

\title[Non-Gaussanity of Diffuse Galactic synchrotron emission]{Non-Gaussianity of diffuse Galactic synchrotron emission at 408 MHz} 

\author[Rana et. al.]
{Sandeep Rana,$^{1}$\thanks{sandeeprana@iisermohali.ac.in}, Tuhin
  Ghosh$^{2}$\thanks{tghosh@niser.ac.in}, Jasjeet
  S. Bagla$^{1}$\thanks{jasjeet@iisermohali.ac.in}, Pravabati
  Chingangbam$^{3}$\thanks{prava@iiap.res.in} \\ 
    $^{1}$Indian Institute of Science Education \& Research, Mohali,
    Knowledge city, Sector 81, S.A.S. Nagar, Manauli, Punjab 140306,
    India\\ 
    $^{2}$School of Physical Sciences, National Institute of Science
    Education and Research, HBNI, Jatni 752050, Odissa, India\\ 
    $^{3}$Indian Institute of Astrophysics, Koramangala II Block,
    Bangalore 560034, India} 

\begin{document}
\date{Accepted . Received ; in original form}
\maketitle

% Abstract of the paper
\begin{abstract}
    Diffuse Galactic emission at low frequencies is a major contaminant
    for studies of redshifted $21$~cm line studies.
    Removal of these foregrounds is essential for exploiting the signal
    from neutral hydrogen at high redshifts.
    Analysis of foregrounds and its characteristics is thus of utmost
    importance. 
    It is customary to test efficacy of foreground removal techniques
    using simulated foregrounds.
    Most simulations assume that the distribution of the foreground
    signal is a Gaussian random field. 
    In this work we test this assumption by computing the binned
    bispectrum for the all-sky $408$~MHz map.
    This is done by applying different brightness temperature ($T$)
    thresholds in order to assess whether the cooler parts of the sky
    have different characteristics.
    We find that regions with a low brightness temperature $T < 25$\,K
    indeed have smaller departures from a Gaussian
    distribution.
    Therefore, these regions of the sky are ideal for future
    \hi~intensity mapping surveys.  

\end{abstract}


%%%%%%%%%%%%%%%%%%%%%%%%%%%%%%%%%%%%%%%%%%%%%%%%%%

%%%%%%%%%%%%%%%%% BODY OF PAPER %%%%%%%%%%%%%%%%%%

\section{Introduction}

Diffuse Galactic emission at low frequencies is of immense interest on
many counts.
These emissions allow us to study the interstellar medium  in
the Galaxy in unprecedented detail.
However, for astronomers interested in radiation from galaxies at high
redshifts, radiation from the interstellar medium of the Galaxy is
a contaminant. 
Further, we find that this component is brighter than the radiation
from high redshift sources by several orders of magnitude
{\citep{Furlanetto}}.
The nature of radiation from high redshift sources is very
different from the continuum radiation from the interstellar medium.
This difference in the spectral properties helps in delineating the
radiation from two different sources.
The methods deployed in such cases can be categorised in three
classes: 
\begin{itemize}
\item
  Foreground avoidance \citep{Dutta, Pober} 
\item
  Foreground removal \citep{Morales, Harker1} 
\item
  Foreground suppression \citep{TGE, CHIPS} .
\end{itemize}
All approaches to disentangling signal of interest from foregrounds
depends on our understanding of the nature of foregrounds. 
Therefore understanding and characterising Galactic foregrounds are
critical aspects of designing next generation radio surveys especially
\hi\ intensity mapping experiments for the epoch of reionisation (\eor) as
well as post-reionisation distribution of \hi, such as Murchison Wide Field 
Array (MWA, \citealt{MWA1}), 
Low-Frequency Array (LOFAR, \citealt{LOFAR1}),
Ooty Wide Field Array (OWFA, \citealt{OWFA1}). 

The Galactic foregrounds consist of many components with the dominant
source at low frequencies being the Galactic diffuse synchrotron
emission.
At higher frequencies, the free-free emission and dust also contribute
significantly and are relevant for studies of the Cosmic Microwave
Background  radiation (CMB).

Various investigations to understand the Galactic  foregrounds for
CMB and \hi\ intensity mapping experiments have laid down the
generic assumptions. 
Using inputs from these, we can model spatial and frequency variation
of Galactic foregrounds in simulations \citep{Wang, Liu11, 
  Jelic08, Jelic10, Santos, Waelkens}. 
The majority of these models use a power-law spectrum with a slowly
varying index and an amplitude as free parameters. 
These Galactic foreground sky simulations assume that parameters of
model are approximated by Gaussian random field (GRF), e.g. see
\cite{Jelic08, Tegmark, Santos, Shaw}.
In this work we analyse the observed emission to test the validity of
the assumption of the angular distribution being close to a Gaussian
random field.
This is done for different subsets of the data where the subsets are
defined using the brightness temperature. 

While it is has been proposed that foregrounds are limited
  to the wedge in Fourier space, it has been found that there is
  some leakage beyond the wedge due to spectral variation in
  foregrounds, especially at small wave numbers
  \citep{Pober}.
  Further, most foreground removal schemes \citep{Morales, Liu09,
    Harker1}  assume that foregrounds vary smoothly in frequency
  and hence can be modelled by a simple function.
  However, observations suggest  \citep{Ghosh} that the spectral
  response may be fairly non-trivial.
  Simulation studies for LOFAR \citep{Chapman} where they test
  foreground removal techniques find that foreground avoidance and
  removal methods work correctly in case of smooth foregrounds,
  but failed in case of small wiggles in foreground amplitude with
  frequency. 
  Thus there is a strong possibility of residuals from foreground
  subtraction introducing higher moments in spatial variation,
  especially if foregrounds are non-Gaussian.
  Thus it is very useful to know if foregrounds have significant
  non-Gaussian features.
  The \hi-signal itself is expected to be slightly non-Gaussian
  due to non-linear aspects of gravitational clustering
  \citep{Somnath}.
  \citet{Cooray} proposed a foreground removal technique where the
  prior knowledge of non-Gaussian structure of foregrounds can be
  combined with its frequency information to separate out
  \hi~fluctuations from the foregrounds. 

A number of methods for quantifying departure from Gaussianity have
been suggested, e.g., Kullback-Leibler divergence \citep[see][]{Dav15b},  
higher order moments (skewness, kurtosis) \citep{Dav15a,
  skew_spectra, psuedo_cl}, and three-point correlation function
\citep{3-pt_corr}. 
One of the tools for analysing the statistical properties of
smooth random fields are the Minkowski
Functionals~\citep{Adler:1981,Tomita:1986}.
They are quantities that characterise the geometrical and
topological properties of excursion sets of random fields.
They contain correlations of arbitrary order and hence are very
useful for searches for non-Gaussianity in observed cosmological
data.
They were first applied to cosmological fields, particularly 
the CMB \citep{Gott:1990,Mecke:1994,Schmalzing:1997, Schmalzing1998}. 
They have been extensively used to constrain primordial non-Gaussianity in
the CMB data~\citep{COBE_NG:2000,WMAP_NG:2011,Planck_Nongauss2016,
  Buchert:2017uup}.   
They have also been used to detect the presence of residual
foreground contamination in cleaned CMB
data~\citep{Chingangbam:2013} and to study the effect of lensing
on CMB fields~\citep{Munshi:2016}.  

\citet{Dav15a} using the all-sky renewed version of Haslam $408$~MHz 
map \citep{Remaz} reported that a patch of $3\pdeg7 \times
3\pdeg7$ at higher Galactic latitudes could be well approximated
by GRF (see Figs.~4 and 5 of \citealt{Dav15a}). 
However their results are more localised as they have obtained
local skewness and kurtosis for specific patches in the sky.
In contrast to their approach, we are using the binned bispectrum
estimator \citep{Bucher10, Bucher15} to test the validity of the
Gaussian approximation. 
In view of the large area surveys envisaged by Square Kilometre
Array (SKA) and it precursors (MWA, LOFAR, and OWFA) we do not
limit ourselves to a small patch and we work with different sky
brightness thresholds. 

Bispectrum estimation is a well studied topic in views of several
attempts to search for primordial non-Gaussianity in the CMB
\citep{Gangui,K2001, Bucher10,bispectrum1, Casapo}.  
Non-Gaussianity can also arise due to the Sachs-Wolfe effect,
gravitational lensing, reionisation,
etc. \citep{second_order_anisotropy1,second_order_anisotropy2,
  second_order_anisotropy3, second_order_anisotropy4}. 
For present work, we use the all-sky 408 MHz map  \citep{Remaz} and 
binned bispectrum estimator \citep{Bucher10, Bucher15} to quantify 
non-Gaussian nature of the Galactic foregrounds.
In Sect. \S{2}, we discuss simulations of Galactic foregrounds,
formalism for binned bispectrum estimator.
In Sect. \S{3} we discuss the analysis and results of bispectum
analysis. 
The Minkowski functional analysis is presented in Sect. \S{4}. 
Summary and discussions are presented in Sect. \S{5}.

%%%%%%%%%%%%%%%%%%%%%%%%%%%%%%%%%%%%%%%%%%%%%%%%%%%%%%%%%%%%%%%%%%%%%%%%%%%%%
%%%%%%%%%%%%%%%%%%%%%%%%%%%%%%%%%%%%%%%%%%%%%%%%%%%%%%%%%%%%%%%%%%%%%%%%%%%%%

\section{Formalism}
\label{sec-form}

The low-frequency Galactic foregrounds consists of three components -
diffuse synchrotron emission, diffuse free-free emission from ionised gas
and emission from point sources, which mainly include stellar remnants.
At low frequencies the free-free is weakest among the three  
components but still a dominant component as compared to
cosmological signal (see Fig.~5 of \citealt{Santos}). 
However at higher frequencies ($\geq 50$~GHz) free-free becomes a 
dominant foreground along with thermal dust.

The intensity of the diffuse Galactic synchrotron emission is expressed 
in terms of its angular power spectrum ($C_{\ell}$) where
$C_{\ell} \propto \ell^{\alpha} \nu^{2\beta}$. 
Here $\alpha$ corresponds to the variation of the power spectra as a 
function of angular separation on the sky and $\beta$ corresponds to the
variation in frequency \citep{Tegmark, Giardino, Oliveira, Liu11}.

To simulate the effect of Galactic foreground one assumes
a power-law form with indices $\alpha$ and $\beta$. 
For low-frequency cosmological observations, the Haslam map 
at 408~MHz acts as standard template for  studying diffuse
Galactic foregrounds \citep{Haslam, Remaz}.  
The spectral index varies in the range $\alpha=-2.5$ to $-3.0$
down to a degree angular scale at $408$~MHz \citep{Tegmark,
  Oliveira}. 

\cite{Tegmark} report frequency spectrum varying as $\beta \sim
-2.8$ with $\Delta \beta \sim 0.15$, that is the variation in
frequency spectral index arises due to variation along different
line of sights.  
Studies of low-frequency foregrounds in the frequency range
$100-200$~MHz find standard deviation in $\beta$ to be approx
$0.1$ \citep{Shaver}.  
This has also been studied elsewhere, e.g., see \citet{Rogers,
  Ali, Ghosh, Pen}.  
In contrast to earlier approach, \citet{Jelic08} use to simulate
$T$ in four-dimensional accounting for the variation of
foregrounds along the line of sight ($z$ coordinate): 
\begin{equation}
  T(x, y, \nu) = C \int A(x, y, z, \nu) dz \ , 
\end{equation}
where $C$ is the normalisation constant and $A(x, y, z, \nu)$ is
modelled as power-law $A(x, y, z, \nu) = A(x, y, z, \nu_0)
\big(\frac{\nu}{\nu_{0}} \big)^{\beta(x, y, z)}$. 

Several simulation studies of foregrounds have been carried out
\citep{Jelic08, Jelic10, Santos, Waelkens}.  
Some of these encompass a broad range of frequencies while others
are more specialised for a specific application.
Most of these simulations model intensity and power-law index of
the diffuse synchrotron emission as GRFs.  

For example, the publicly available \hammurabi\ code, simulates
all-sky maps of the polarised synchrotron emission, free-free
emission and ultra-high energy cosmic ray at all frequencies in a
three-dimensional (3D) \healpix\ grid format \citep{Waelkens}.  
It assumes a 3D model of the Galactic  magnetic field, cosmic ray
electron density and thermal density distribution, for the
detailed description see \citet{Waelkens}.  
The turbulent component of the magnetic field is simulated as a
Gaussian random-field realisation given a magnetic-field power
spectrum. 
The realistic turbulent field is not Gaussian, therefore the
simulations fail to reproduce the higher order statistics of the
data. 
The LOFAR \eor\ simulation pipeline \citep{Jelic08,Jelic10} also
assumes that both $A(x, y, z, \nu)$ and the power-law index
$\beta(x, y, z, \nu)$ of diffuse Galactic synchrotron emission as
a GRFs. 

The aim of the present study is to evaluate the validity of the
use of GRFs to represent the $408$~MHz brightness temperature
fluctuations on the sky.  
We do this by computing the magnitude of departure of the observed
spatial distribution from a Gaussian distribution.  
Of course, projection effects in models that take spatial distributions 
in 3D may introduce departures from the Gaussian distribution.

In the present analysis, we are only considering synchrotron
intensity map and will not comment on the polarisation of diffuse
synchrotron emission as this is beyond the scope of this work.  
We quantify non-Gaussianity of the diffuse synchrotron emission by
computing the three-point correlation function in harmonic space
as a function of sky brightness.  

%%%%%%%%%%%%%%%%%%%%%%%%%%%%%%%%%%%%%%%%%%%%%%%%%%%%%%%%%%%%%%%%%%%%%%%%%%%%% 
%%%%%%%%%%%%%%%%%%%%%%%%%%%%%%%%%%%%%%%%%%%%%%%%%%%%%%%%%%%%%%%%%%%%%%%%%%%%% 

\subsection{Binned bispectrum estimator}
\label{subs-Bis}

Bispectrum estimation is a popular approach for quantifying
non-Gaussianity  \citep{Gangui,K2001, bispectrum5, KSW,
  bispectrum1}.
It is particularly used for non-Gaussianity studies of CMB
anisotropies in temperature and polarisation
\citep{Planck_Nongauss2016}.  

Given a distribution of brightness temperature $T$ over all-sky, one can 
decompose it into spherical harmonic basis with amplitudes $a_{\ell m}$
\begin{equation}
  T(\Omega)= \sum_{\substack{\ell,m}}  a_{\ell m}  Y_{\ell m}(\Omega) \ .
\end{equation}
We can compute amplitudes $a_{\ell m}$ from the temperature
distribution on the sky as 
\begin{equation}
  a_{\ell m} = \int T(\Omega) \, Y_{\ell m}^{\ast}(\Omega) \, d\Omega \ .
\end{equation}
The angle-averaged angular bispectrum is defined as
\begin{equation}
  B_{\ell_{1}\ell_{2}\ell_{3}} =  N \sum_{\substack{m_{1}, m_{2},
      m_{3}}} \left( \begin{array}{ccc} 
                       \ell_{1} &\ell_{2} &\ell_{3} \\
                       m_{1} & m_{2} & m_{3} \end{array} \right) a_{\ell_{1}m_{1}}
                   a_{\ell_{2} m_{2}} a_{\ell_{3} m_{3}} \,
\end{equation}
where 
\begin{equation}
  N
  =\frac{(2\ell_{1}+1)(2\ell_{2}+1)(2\ell_{3}+1)
  }{4\pi}\left( \begin{array}{ccc}   
                  \ell_{1} &\ell_{2} &\ell_{3} \\
                  0 & 0 & 0 \end{array} \right)^{2} \,
\end{equation}
is the number of possible triangles of side
$(\ell_{1},\ell_{2},\ell_{3})$ on a sphere (see \citealt{wig3j}).
The angle-average bispectrum $B_{\ell_{1}\ell_{2}\ell_{3}}$ is
pairwise symmetric in $(\ell_{1}, \ell_{2}, \ell_{3})$. 
Therefore in our analysis we only consider the subspace $ \ell_{1}
\leq \ell_{2} \leq \ell_{3}$.  
The presence of Wigner $3j$ symbol imposes condition that
$B_{\ell_{1}\ell_{2}\ell_{3}} \ne 0$ only for $m_{1}+m_{2}+m_{3}=
0$ and $| \ell_{1} - \ell_{2} | \leq \ell_{3}  \leq \ell_{1} +
\ell_{2}$.  
Additional selection condition  $\ell_{1} + \ell_{2} + \ell_{3} =
\mathrm{even}$\  comes due to presence of the Wigner $3j$ symbol
when all $m$ values are zero (see \citealt{Gangui, K2001}), which
means that only even parity modes are considered.

The binned bispectrum introduced by \citet{Bucher10} is defined in
terms of maximally filtered brightness temperature as 
\begin{equation}
  b_{\rm i_{1}i_{2}i_{3}} = \frac{1}{C_{\rm i_{1}i_{2}i_{3}}}\int
  d\Omega \  T_{\rm i_{1}} (\Omega) \  T_{\rm i_{2}} (\Omega) \
  T_{\rm i_{3}} (\Omega) \ , \label{eq:2.1.8} 
\end{equation}
where
\begin{equation}
  T_{\rm i} (\Omega) = \sum_{\ell, m} f_{\ell}^{\rm i} \ a_{\ell
    m}\ Y_{\ell m}(\Omega) \ .   \label{eq:2.1.9} 
\end{equation}
$C_{\rm i_{1}i_{2}i_{3}}$ is the number of valid $\ell$ triplets
within $(\rm i_{1},i_{2}, i_{3})$ and $f_{\ell}$ is a filter
function in harmonic space. 
The binned bispectrum estimator is optimised by working with
$T_{\rm i}$, where each value of $\rm i$ refers to a range of
$\ell$ values. 
This scheme of working with bins is described in detail by 
\citet{Bucher10}.
The advantage of using binned bispectrum estimator is that it is
computationally faster than the angle-averaged bispectrum
estimator.  
The binned bispectrum $b_{\rm i_{1}i_{2}i_{3}}$ is related to
angle-averaged bispectrum by the relation
\begin{equation}
  b_{\rm i_{1}i_{2} i_{3}} =  \sum_{\ell_1 \ell_2 \ell_3}
  f_{\ell}^{\rm i_1} f_{\ell}^{\rm i_2} f_{\ell}^{\rm i_3}
  \frac{B_{\ell_{1}\ell_{2}\ell_{3}} }{\sqrt{N}} \ . 
\end{equation}
Binned bispectrum, by construction, includes both odd and
  even parity modes, and there is no ab initio reason for giving
  preference to specific parity mode in case of the diffused
  Galactic synchrotron emission.  
  Therefore in our analysis we include both parity modes.

In presence of partial sky coverage and a finite beam resolution
of a given map, the binned bispectrum estimator defined in
Eq.~\ref{eq:2.1.8} needs to be debiased.
The linear order correction term is written as \citep{Bucher15}
\begin{equation}
  b_{\rm i_{1}i_{2}i_{3}}^{\rm lin} = \frac{1}{C_{\rm
      i_{1}i_{2}i_{3}}} \int d\Omega \left[ 
    T_{\rm i_{1}} \langle  T_{\rm i_{2}}^{\rm G} T_{\rm
      i_{3}}^{\rm G} \rangle   
    +T_{\rm i_{2}} \langle  T_{\rm i_{1}}^{\rm G} T_{\rm
      i_{3}}^{\rm G} \rangle  
    +T_{\rm i_{3}} \langle  T_{\rm i_{1}}^{\rm G} T_{\rm
      i_{2}}^{\rm G} \rangle  
  \right] \ .
\end{equation}
Here the term inside angular brackets is an average over 1000 Gaussian
realisations which have the same underlying power spectrum as the
observed data. 
We discuss this in detail in the next section.

%%%%%%%%%%%%%%%%%%%%%%%%%%%%%%%%%%%%%%%%%%%%%%%%%%%%%%%%%%%%%%%%%%%%%% 

\section{Analysis}

\subsection{Masking}\label{sec:3.1}

To investigate the non-Gaussian nature of diffuse synchrotron
emission, we use a renewed version of all-sky 408MHz map, which is
further de-sourced and de-striped to remove any remaining point
source contamination that was present in original Haslam map
\citep{Remaz}. 

Taking this as a reference map, we apply sky brightness thresholds
to take out prominent non-Gaussian Galactic structures.
We have used the publicly available
\healpy\footnote{https://pypi.python.org/packages/source/h/healpy/healpy-1.7.4.tar.gz}
package, the python version of
\healpix\footnote{http://healpix.sourceforge.net} \citep{Healpix},
and employ the following steps:   
\begin{itemize}
\item
  First, we degrade $408$~MHz map from \Nside$=512$ to $128$ since
  the resolution of processed $408$~MHz map is $56\arcmin$
  \citep{Remaz}.
  We then apply a Galactic cut ($|b| < 10\deg$) to take out
  contamination from the Galactic plane. 
\item
  We mask bright Loop-I ring, which comes from an old supernovae
  remnant.
  The mask region includes $\pm 4\deg$ width cut around a circle
  of radius $58\deg$ entered at $(l, b)=(329\deg, 17\pdeg5)$. 
\item
  We construct four sky masks by applying brightness temperature
  thresholds of $25$, $30$, $40$, and $60$~K.  
\end{itemize}

\begin{figure*}
  \includegraphics[width=14.5cm]{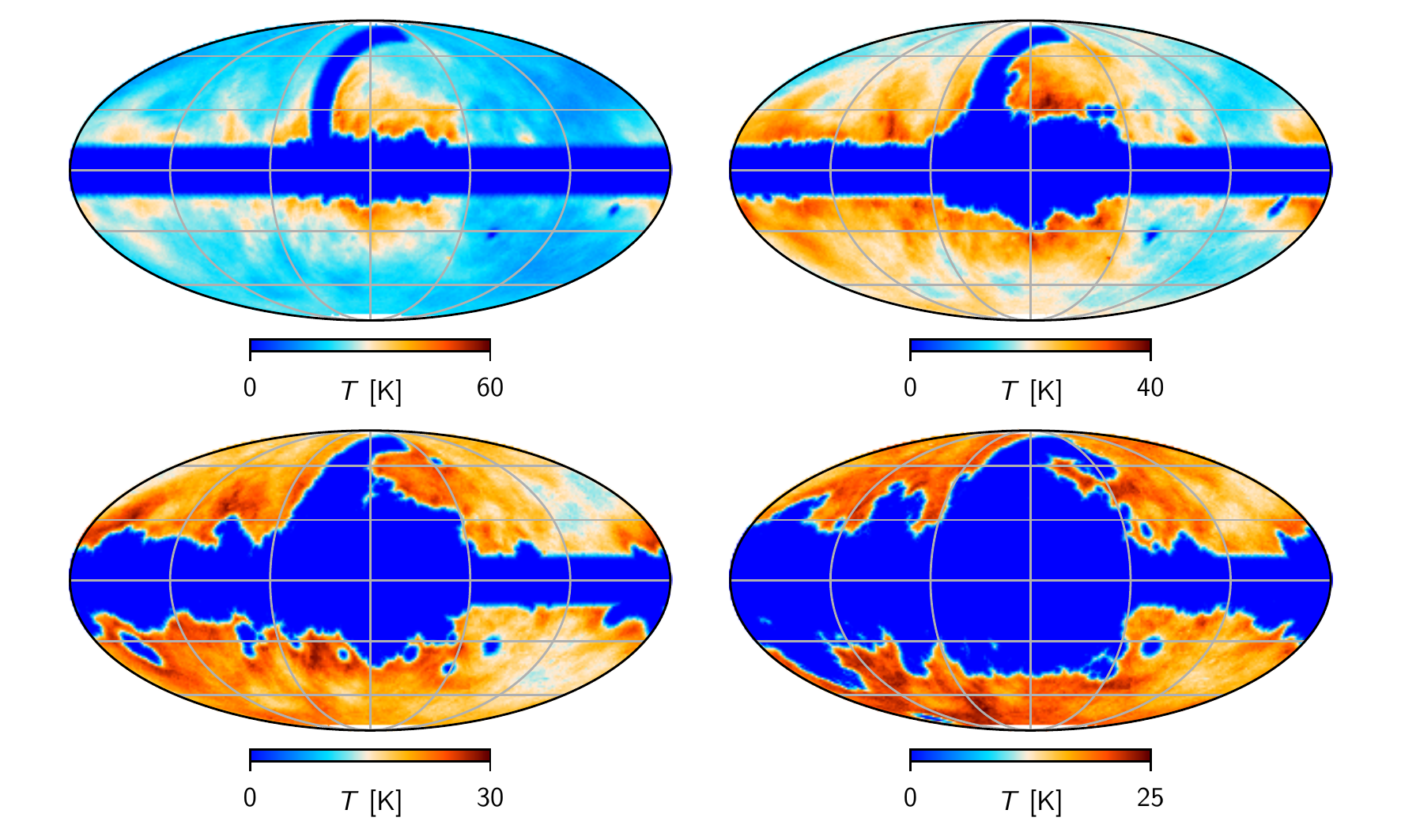}
  \caption {408 MHz map multiplied by a $5\deg$ FWHM Gaussian
    apodized masks at different sky brightness thresholds as
    described in Sect.~\ref{sec:3.1}.
    Here temperature scales are different for each map to show
    brightness temperature variation.
    The effective fractional sky coverages, $f_{sky}$, are $0.74$
    ($T < 60$~K), $0.63$ ($T < 40$~K), $0.56$~($T < 30$~K), and
    $0.41$ ($T < 25$~K) respectively.} 
  \label{1}
\end{figure*}

To avoid leakage of power due to sharp cutoff between the masked
and the unmasked region, we apodized the masks by convolving with a
$5\deg$ FWHM Gaussian.  
Such a smooth Gaussian filter function reduces the leakage of the
signal towards the edges of the unmasked region.
Figure~\ref{1} shows the $408$~MHz map multiplied by the apodized
masks at different sky brightness thresholds.
The effective fractional sky coverages, $f_{\rm sky}$, are $0.74$
($T < 60$~K), $0.63$ ($T < 40$~K), $0.56$ ($T < 30$~K), and $0.41$
($T < 25$~K).  

%%%%%%%%%%%%%%%%%%%%%%%%%%%%%%%%%%%%%%%%%%%%%%%%%%%%%%%%%%%%%%%%%%%%%%%%%%% 

\subsection{Bispectrum estimation} \label{sec:3.2}

After apodization of masks, we use
\polspice\footnote{http://www2.iap.fr/users/hivon/software/PolSpice/index.html}
routine \citep{PolSpice} to compute the full-sky $TT$ power
spectra at different sky brightness.
The \polspice\ routine corrects for the sky mask, beam smoothing
and pixel window function.
The correction for incomplete sky coverage is performed by first
estimating the correlation function in pixel space.  
As $408$~MHz map is non-Gaussian and anisotropic, all the
statistical properties of the Galactic synchrotron emission might
not be captured by the power spectrum analysis. 
For further analysis, we take each of the four $TT$ power spectrum
and simulate $1000$ Gaussian sky realisations using \healpy\
facility \synfast\ and apply the Gaussian beam of FWHM $56\arcm$.
We compare the two-point statistics of the data and the Gaussian
simulations for multipole range where they are reliably
estimated.
The lower bound on $\ell$ value comes from the masking of the sky
region.
The resolution of the $408$~MHz map puts the bound on high $\ell$
mode.
Therefore for the power spectrum and bispectrum analyses, we only
consider the $\ell$-range $\in [10, 180]$.

We consider a simple linear binning scheme and divide the
available $\ell$ range into $N_{\rm bin}=11$, where $N_{\rm bin}$
is the total number of bins.
The binning scheme is defined as follows:
\begin{equation}
  f_{\ell}^{\rm i}= \begin{cases}
    \cos^{2}\left(\frac{\pi}{2} \times \frac{\ell_0^{\rm i} -
        \ell}{\Delta \ell}\right) & \text{for $\ell_{0}^{\rm i} -
      \Delta \ell \leq \ell < \ell_{0}^{\rm i}$} \\ 
    1  & \text{for $\ell_0^{\rm i} \leq \ell
      \leq\ell_1^{\rm i}$} \\ 
    \cos^{2}\left(\frac{\pi}{2} \times \frac{\ell- \ell_{1}^{\rm
          i} }{\Delta \ell}\right) & \text{for $\ell_{1}^{\rm i} <
      l \le \ell_{1}^{\rm i} +\Delta \ell$ } \ , 
  \end{cases}
\end{equation}
where $\ell_0^{\rm i}$ and $\ell_1^{\rm i}$ are the first and the
last $\ell$ values of a given $\rm i$ bin and $\Delta \ell = 5$. 
Table~\ref{tab:1} summarises the binning scheme that we employ in
our analysis. 
The filter function is defined in such a way that $\sum_{\rm i}
(f_{\ell}^{\rm i})^2=1$ for all bins, except for the first and
last bins. 
It has been shown that the variance of the binned bispectrum
changes with the binning scheme.
Therefore, one has to optimise the binning scheme to obtain the
minimum variance binned bispectrum estimator \citep{Bucher10,
  Casapo, Bucher15}.
In the present analysis, we only consider one binning scheme.
We have checked that the final results of the paper does not
depend on the binning scheme. 

\begin{figure}
  \includegraphics[width=8.8cm]{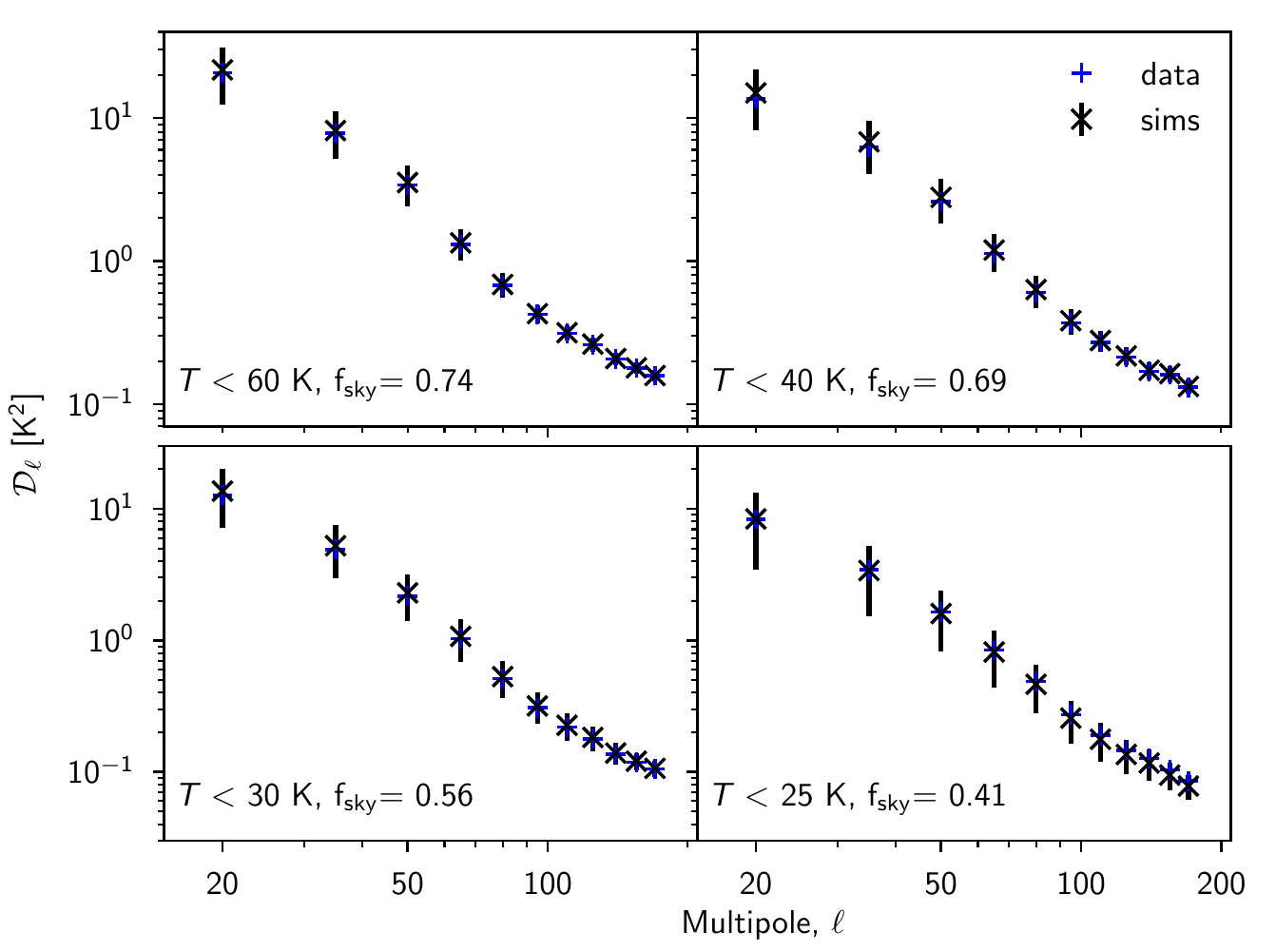}
  \caption {Binned angular power spectrum as a function of
    multipoles for different sky brightness thresholds, blue
    circles is for real data and black open circles is a mean of
    $1000$ Gaussian realisations.
    The data agrees with the Gaussian realisations within
    $1\sigma$ confidence limit.} 
  \label{2}
\end{figure}

Once the binning scheme and $\ell$-range are defined, we then
calculate $D_{\ell}=\ell (\ell+1)C_{\ell}/2\pi$ in each $\ell$ bin
for all $1000$ Gaussian realisations and evaluate the ensemble
mean and standard deviation value of $D_{\ell}$. 
Similarly, we obtain $D_{\ell}$ in each $\ell$ bin for the real
data.
In Fig.~\ref{2} we plot both averaged and actual binned
$D_{\ell}$, and both of them agree well within 1$\sigma$ and
ensures that simulated maps are in good agreement with actual
masked $408$~MHz map as far as the power spectrum is concerned. 

As a first step of bispectrum estimation, we create binned
filtered maps as given by Eq.~\eqref{eq:2.1.9} in each $\rm i$
bin.
Upon close inspection of these filtered maps, we have realised
that the bright Galactic synchrotron emission from the Galactic
plane leaks into the high latitude sky.
To avoid this problem, we first mask the all-sky $408$ MHz map
with a sky brightness threshold of $80$~K and then compute the
binned filtered maps. 
With $80$~K threshold, we only take out the brightest part of the
synchrotron emission close to the Galactic plane.
We then apply the four binary masks to the filtered $408$~MHz
maps.
The same procedure is applied to the $1000$ Gaussian sky
realisations for each sky brightness threshold. 

The numerical implementation of the unbiased binned bispectrum estimator is 
\begin{align}
  b_{\rm i_{1}i_{2}i_{3}}^{\rm o} & = \frac{1}{C_{\rm
                                    i_{1}i_{2}i_{3}}}
                                    \sum_{j=1}^{N_{\rm pix}}
                                        \frac{M_{\rm j} }{4\pi \Npix}
                                    ( T^{\rm i_{1}}_{\rm j} T^{\rm
                                    i_{2}}_{\rm j} T^{\rm
                                    i_{3}}_{\rm j}   \nonumber \\  
                                  & -  T^{\rm i_{1}}_{\rm j} \langle T^{\rm i_{2}}_{\rm j} T^{\rm 
                                    i_{3}}_{\rm j} \rangle + T^{\rm i_{2}}_{\rm j} \langle T^{\rm
                                    i_{1}}_{\rm j} T^{\rm i_{3}}_{\rm j} \rangle + T^{\rm
                                    i_{3}}_{\rm j} \langle T^{\rm i_{1}}_{\rm j} T^{\rm
         i_{2}}_{\rm j} \rangle ) \ , 
\end{align}
where $M$ is the binary mask applied binned filtered maps,
$\Npix=\sum_{\rm j} M_{\rm j}$ is the fraction of sky available,
$C_{\rm i_{1}i_{2}i_{3}}$ is the number of valid triplets within
the bin ${\rm i_{1}i_{2}i_{3}}$.  
If there is no valid triplet in $(\rm i_{1},i_{2},i_{3})$ bins, we
put  $C_{\rm i_{1}i_{2}i_{3}}=0$ and ignore the contribution to
binned bispectrum.
We apply the above estimator to obtained binned bispectrum value
for the four sky masks.
Similarly, we repeat the same procedure on $1000$ Gaussian sky
realisations that have the same underlying power spectrum as the
observed $408$~MHz map.
From the simulations, we obtain the ensemble mean and $1\sigma$
standard deviation of the binned bispectrum value as a function of
sky brightness.

%%%%%%%%%%%%%%%%%%%%%%%%%%%%%%%%%%%%%%%%%%%%%%%%%%%%%%%%%%%%%%%%%%%%%%%%%% 
    
\subsection{Bispectrum visualisation}

To establish the statistical significance of the observed binned
bispectrum of $408$~MHz map as a function of sky brightness
thresholds, we compute the deviation, $\Delta = (b_{\rm
  i_{1}i_{2}i_{3}}^{\rm o} - \langle b_{\rm i_{1}i_{2}i_{3}}^{\rm
  G}\rangle)/\sigma_{\rm i_{1}i_{2}i_{3}}^{\rm G}$, between the
data and the ensemble mean of large set of $1000$ Gaussian
simulations with the same underlying angular power spectrum as the
$408$~MHz map.
The $1\sigma$ standard deviation of $1000$ Gaussian simulations
binned bispectrum is denoted by the index $\sigma_{\rm
  i_{1}i_{2}i_{3}}$.
Unlike the primordial non-Gaussianity study of CMB, we do not have
a standard bispectrum template to compute local non-linear
  coupling parameter ($f_{\rm NL}$). 

\begin{figure}
  \begin{center}
    \includegraphics[width=9.0cm,angle=0.0]{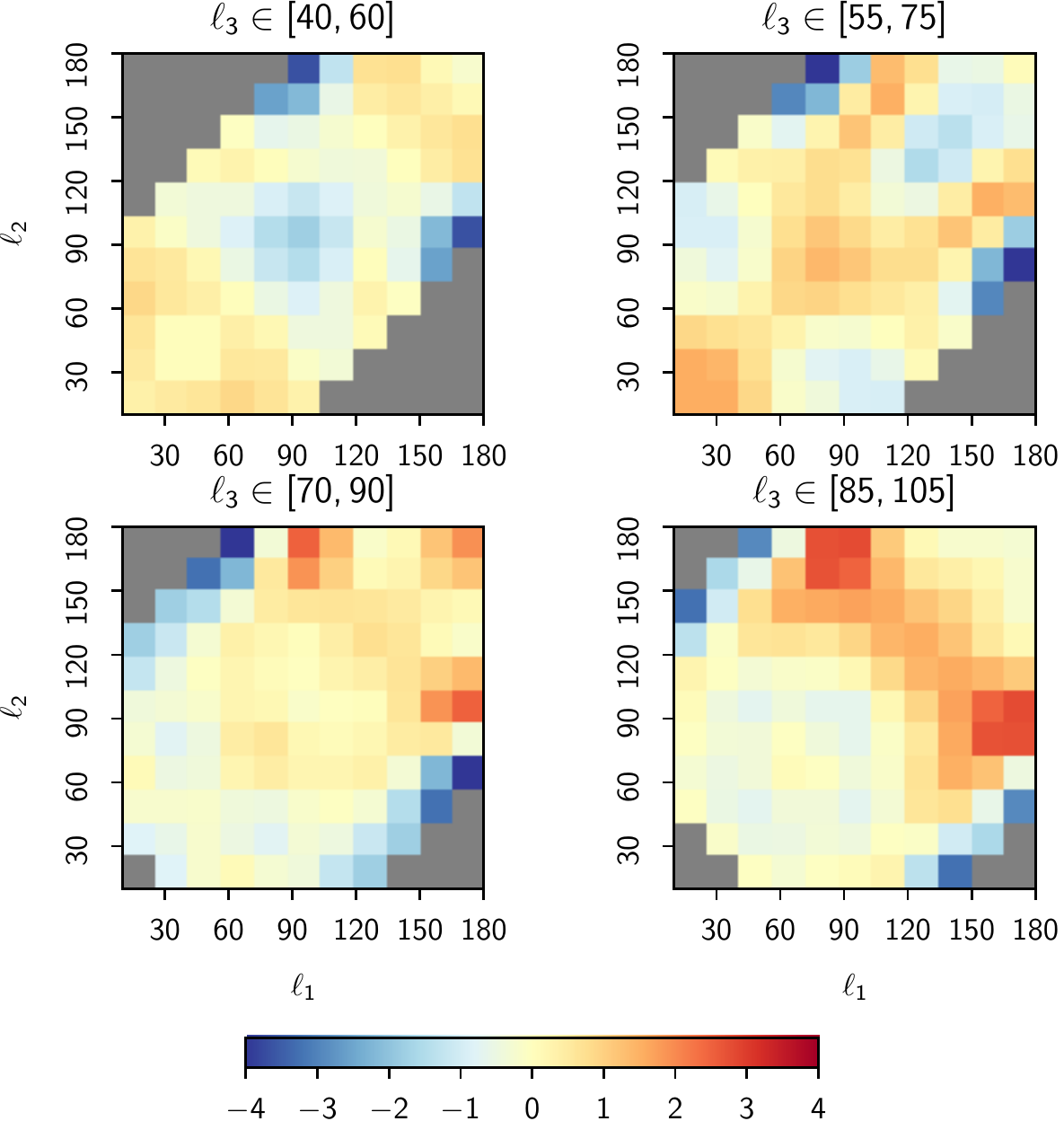}
    \caption{Variation of $\Delta$ in $\ell_{1}-\ell_{2}$ plane fixing
      $\ell_{3}$ bin range for 408~MHz brightness threshold
      $T~<~60$\,K.
      The plots are symmetric $\ell_1-\ell_2$ plane because of the
      symmetric nature of $b_{\ell_{1}\ell_{2}\ell_{3}}$ under the
      permutation of $\ell$ pairs.} 
    \label{3}
  \end{center}
\end{figure}

\begin{figure}
  \begin{tabular}{c}
    \includegraphics[width=9.0cm,angle=0.0]{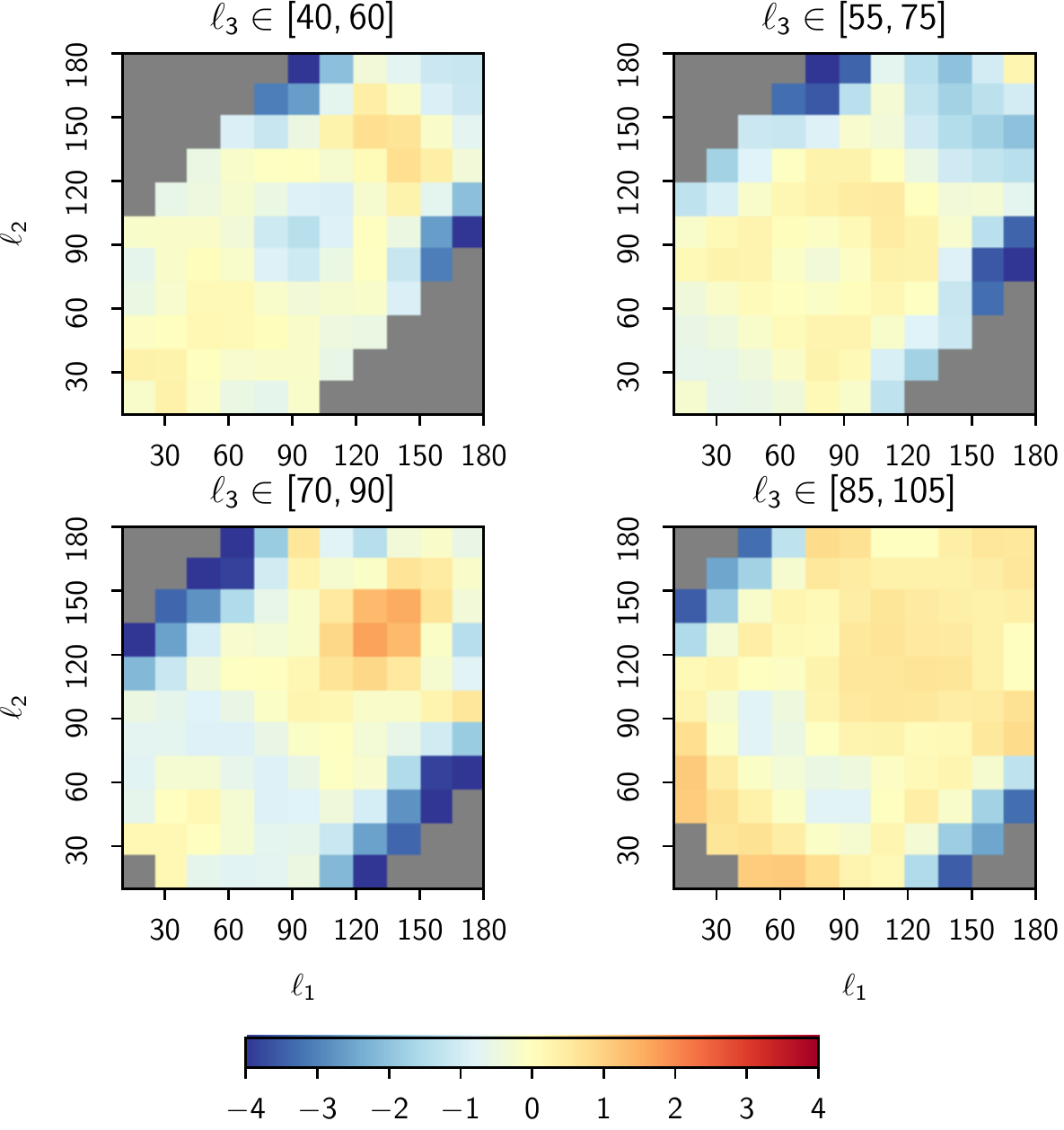}
  \end{tabular}
  \caption{Same as Fig.~\ref{3}, but for 408~MHz brightness threshold
    $T~<~40$\,K.} 
  \label{4}
\end{figure}

\begin{figure}
  \begin{tabular}{c}
    \includegraphics[width=9.0cm,angle=0.0]{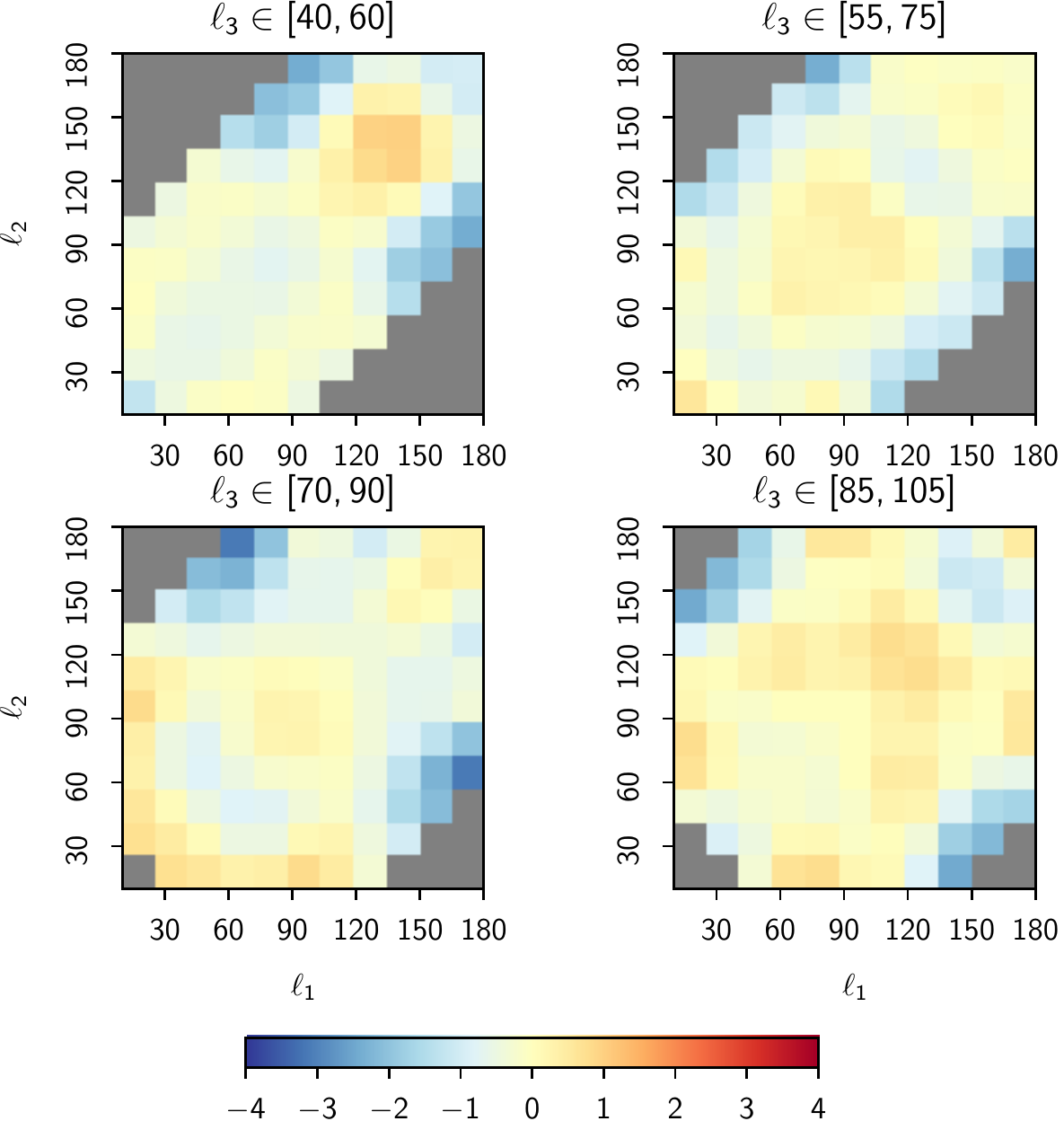}
  \end{tabular}
  \caption{Same as Fig.~\ref{3}, but for 408~MHz brightness threshold $T~<~30$\,K.}
  \label{5}
\end{figure}

\begin{figure}
  \begin{tabular}{c}
    \includegraphics[width=9.0cm,angle=0.0]{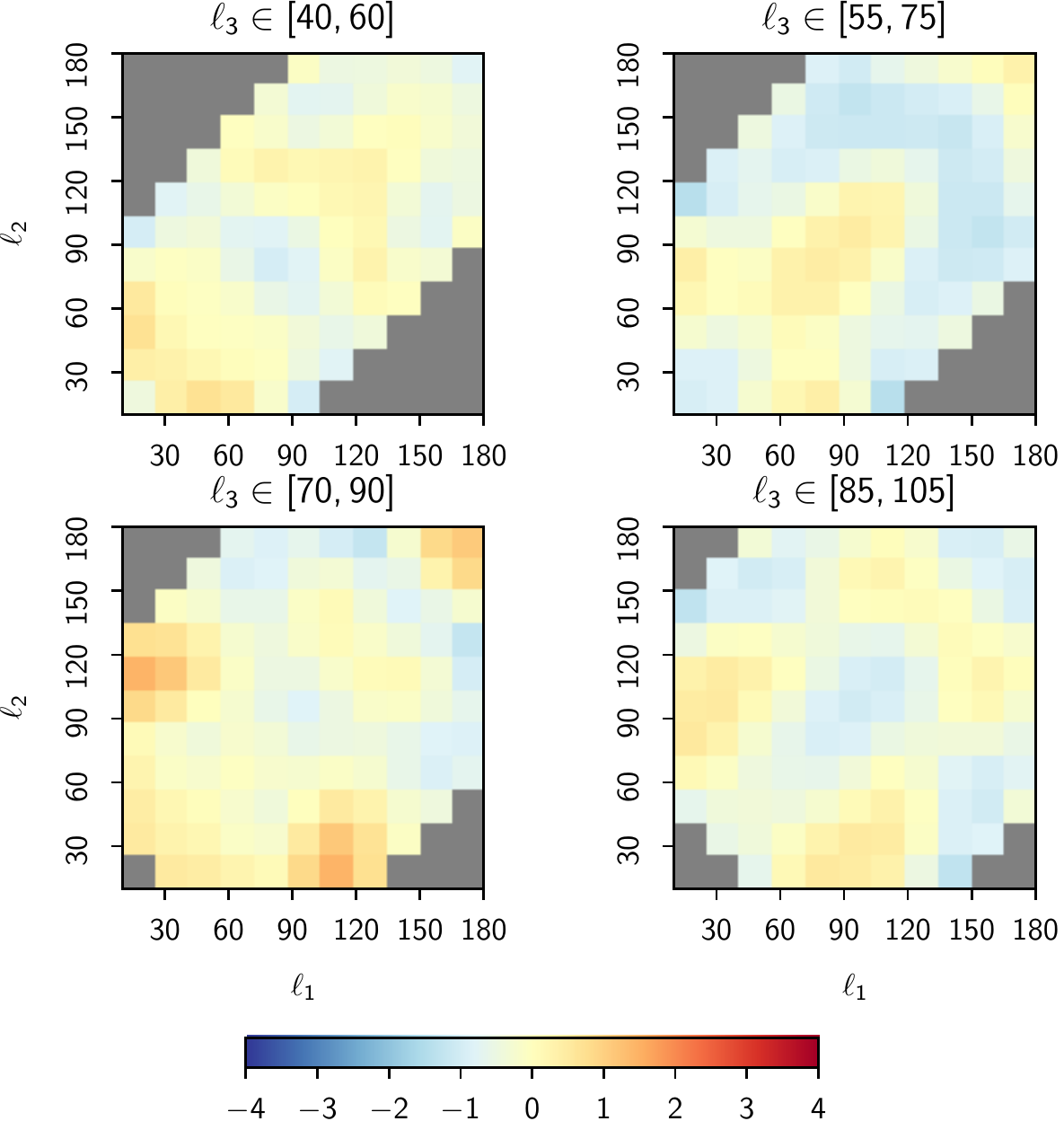}
  \end{tabular}
  \caption{Same as Fig.~\ref{3}, but for 408~MHz sky brightness
    threshold $T~<~25$\,K.} 
  \label{6}
\end{figure}

Figures~\ref{3},  \ref{4}, \ref{5} and \ref{6} show the
two-dimensional plot of $\Delta$ in $\ell_1-\ell_2$ plane fixing
$\ell_3$ parameter. 
The plots are symmetric about the diagonal in $\ell_1-\ell_2$
plane because of the symmetric nature of $b_{\rm
  i_{1}i_{2}i_{3}}^{\rm o}$ under the permutation of $\ell$
pairs.

\begin{figure}
  \begin{tabular}{c}
    \includegraphics[width=9.0cm,angle=0.0]{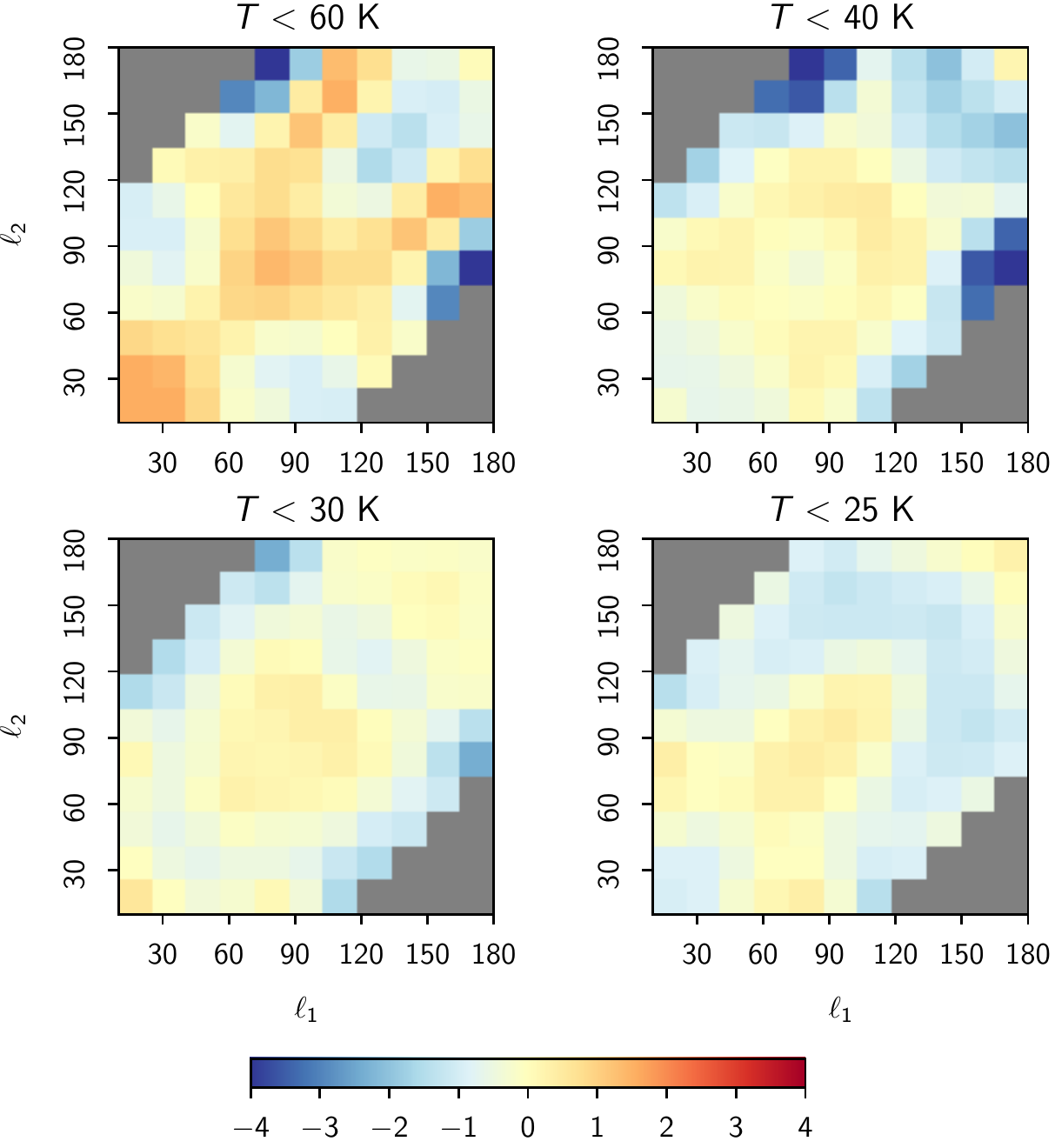}
  \end{tabular}
  \caption{Variation of $\Delta$ 
    plane by fixing $\ell_{3}\in[55, 75]$ for different sky brightness
    thresholds.
    It clearly shows decreasing trend of 
    bispectrum signal as we go from high to low $T$ maps.}
  \label{7}
\end{figure}

Similarly in Fig.~\ref{7}, we plot $\ell_{3}\in[55, 75]$ for
different sky brightness thresholds. 
One can see a clear decreasing trend in bispectrum signal as we go from 
$60$~K maps to $25$~K thresholds on all angular scales.
Table~\ref{2} shows the $(\rm i_{1},i_{2},i_{3})$ bin triplets
where the absolute value of the bispectrum signal of $408$~MHz map
deviate more than $4\sigma$ with respect to ensemble mean of
$1000$ Gaussian realisations. 
As indicated in Table~\ref{2} and Fig.~\ref{6} that in case 
of $25$~K sky brightness threshold, the real data is consistent
within $3\sigma$ deviation with respect to Gaussian
  simulations at $\ell \geq 60$. At large angular scales, $\ell <
  60$, the deviation of the real data varies between 4 and
  $9\sigma$ with respect to the Gaussian simulations. 
  We also compute the skewness and kurtosis of the real and
  imaginary part of $a_{\ell m}$ distribution in $\rm i$ bins for
  sky brightness threshold maps and corresponding 1000 Gaussian
  simulations.  
  We report that both skewness and kurtosis for actual data is in
  agreement with the Gaussian simulations well within $2\sigma$.  
  However, \citet{Pietrobon} show that unlike bispectrum the
  skewness of $a_{\ell m}$ distribution is sensitive only to a
  specific type of triangle configuration and does not represent
  the overall picture as far as non-Gaussian features are
  concerned (see Appendix~\ref{appendix:A}). 

\begin{table}
  \caption {$\ell$ range corresponding to $\rm i$ bin} \label{tab:1} 
  \begin{center}
    \begin{tabular}{c | c | c | c | c} 
      \hline
      bin & $\ell_{min}$ &$\ell_{max}$ & $\ell_0$ & $\ell_1$ \\
      \hline
      $\rm i_{1}$  & 10 & 30 & 15 & 25\\
      $\rm i_{2}$  & 25 & 45 & 30 & 40\\
      $\rm i_{3}$  & 40 & 60 & 45 & 55\\
      $\rm i_{4}$  & 55 & 75 & 60 & 70\\
      $\rm i_{5}$  & 70 & 90 & 75 & 85\\
      $\rm i_{6}$  & 85 & 105 & 90 & 100\\
      $\rm i_{7}$  & 100 & 120 & 105 & 115\\
      $\rm i_{8}$  & 115 & 135 & 120 & 130\\
      $\rm i_{9}$  & 130 & 150 & 135 & 145\\
      $\rm i_{10}$ & 145 & 165 & 150 & 160\\
      $\rm i_{11}$ & 160 & 180 & 165 & 175\\
      \hline
    \end{tabular}
  \end{center}
\end{table}

\begin{center}
  \begin{table*}
    \caption { Triplet bins for which $|\Delta| >4$ where, $\Delta
      = (b_{\rm i_{1}i_{2}i_{3}}^{\rm o} - \langle b_{\rm
        i_{1}i_{2}i_{3}}^{\rm G}\rangle)/\sigma_{\rm
        i_{1}i_{2}i_{3}}^{\rm G}$} \label{tab:2}  
    \begin{tabular}{c | c |  c | c | c |  c | c | c | c | c | c }
      \hline
      &   &  & \multicolumn{5}{c}{$T <$ 60 K} & &  \\
      \hline
      (1, 1, 6)&(1, 1, 7)&(1, 1, 9)&(1, 2, 3)&(1, 2, 7)&(1, 2, 8)&(1, 3, 4)&(1, 3, 8)&(1, 4, 5)&(1, 4, 10)&(1, 5, 10)\\
      (1, 6, 7)&(1, 6, 11)&(1, 7, 8)&(1, 8, 9)&(1, 9, 10)&(2, 2, 4)&(2, 3, 9)&(2, 3, 10)&(2, 4, 11)&(2, 5, 7)&(2, 5, 10)\\
      (2, 5, 11)&(2, 6, 11)&(2, 8, 10)&(3, 4, 11)&(3, 7, 10)&(3, 8, 11)&(4, 6, 10)&(4, 7, 10)&(4, 7, 11)&(5, 5, 5)&(5, 5, 10)\\
      (5, 6, 10)&(5, 6, 11)&(5, 10, 10)&(6, 6, 11)&(7, 7, 10)&(7, 8, 9)&(7, 9, 11)\\
      \hline
      &   &  & \multicolumn{5}{c}{$T <$ 40 K} & &  \\
      \hline
      (1, 1, 2)&(1, 1, 7)&(1, 2, 3) & (1, 2, 7) & (1, 3, 4) & (1, 3, 9)&(1, 3, 10)&(1, 4, 8)&(1, 4, 10)&(1, 4, 11)&(1, 5, 6)\\
      (1, 5, 11)&(1, 6, 11)&(1, 7, 8)&(1, 8, 9)&(1, 9, 10)&(1, 10, 11)&(2, 2, 4)&(2, 2, 11)&(2, 3, 9)&(2, 3, 10)&(2, 5, 11)\\
      (2, 6, 11)&(2, 8, 8)&(2, 10, 10)&(3, 4, 11)&(3, 6, 9)&(3, 8, 9)&(4, 6, 10)&(5, 5, 10)&(5, 6, 11)&(7, 10, 10)&\\
      \hline
      &   &  & \multicolumn{5}{c}{$T  <$ 30 K} & &  \\
      \hline
      (1, 4, 9) & (1, 5, 10)  & (1, 7, 8)  & (1, 8, 10) & (2, 6,
                                                          11)  &
                                                                 (3,
                                                                 4,
                                                                 10)
               & (3, 4, 11)  &  (3, 5, 11)  &  (3, 8, 9)  &  (4,
                                                            4, 11)
                                                       &  (5, 6,
                                                         11)  \\    
      \hline
      &   &  & \multicolumn{5}{c}{$T <$ 25 K} & &  \\
      \hline
      (1, 2, 9) & (1, 2, 10)  & (1, 3, 10)  & (1, 4, 8)& (1, 5, 7)
                                                & (1, 6, 10) & (1,
                                                               7,
                                                               11)&
                                                                    (1,8,9)
                                   & (2, 3, 11)  &  (2, 6, 11) &
                                                                 (3,
                                                                 5,
                                                                 11)
      \\  
      \hline
    \end{tabular}
  \end{table*}
\end{center}

%%%%%%%%%%%%%%%%%%%%%%%%%%%%%%%%%%%%%%%%%%%%%%%%%%%%%%%%%%%%
\section{Minkowski Functionals}

In CMB studies, the scalar Minkowski Functionals (MFs hereafter) is
widely used to measure the non-Gaussianity \citep{Ducout2013}.
Unlike binned bispectrum estimator, MFs are defined in the pixel
space.
MFs are sensitive to the contribution of higher order correlations,
where binned bispectrum estimator is only sensitive to three-point
correlation function.
We employ MFs estimator to cross validate our findings for bispectrum
in case of $408$~MHz map. 

\subsection{Method}

The Minkowski Functionals are a class of morphological descriptors for
structures in $d$-dimensional space.
In two dimensions each level or excursion set of a smooth random field
at a chosen threshold field value typically consists of two kinds of
structures, namely, connected regions and holes within the connected
regions.  
The morphological properties of these structures vary smoothly with
the threshold value.  
The functional form of these variations can reveal the statistical
nature of the field.  
For our purpose in this paper we focus on fields defined on the
sphere, $\mathcal S^2$. 

Let $f$ denote a generic random field with its mean and rms denoted by
$\mu$ and $\sigma_0$, respectively. We define $u$ to be the standard
normal field given by $u\equiv (f-\mu)/\sigma_0$.  
Let $\nu$ be the threshold value chosen from the range of $u$, $Q$ be
the excursion set indexed by $\nu$, and $\partial Q$ the set of the
corresponding boundary contours.  
Then, the MFs for the structures of the excursion set are defined  as
follows: 
\begin{eqnarray}
  V_0(\nu) &\equiv& \frac1{4\pi}\int_Q d a, \label{eqn:vo}\\
  \ V_1(\nu) &\equiv & \frac14\int_{\partial Q} dl, \label{eqn:v1}\\
  V_2(\nu) &\equiv& \frac{1}{2\pi}\int_{\partial Q} \kappa \, dl, \label{eqn:v2}
\end{eqnarray}
where $ d a$ is the area element of the excursion set, $\rm dl$ is the
line element on $\partial Q$ and $\kappa$ is the curvature at each
point of the boundary contours.  
It is clear that $V_0$ is the area fraction occupied by the excursion
set on the sphere whose radius is normalised to one.
$V_1$ is related to the total length of the boundary contours, and
$V_2$ is the number of connected regions minus the number of holes.
For an isotropic GRF, analytic expressions for the ensemble
expectation values of the MFs have been derived by
~\citet{Tomita:1986}.
They are:
\begin{eqnarray}
  V_0  &=& \frac12 \, {\rm erfc}\,\left(\frac{\nu}{\sqrt{2}} \right) ,\\
  V_1  &=&  \frac18 \,\left( \frac{\sigma_1}{\sigma_0}\right)\, e^{-\nu^2/2},\\
  V_2  &=& \frac{1}{2\pi^3} \left( \frac{\sigma_1}{\sigma_0} \right)^2
           \, \nu\,e^{-\nu^2/2}  ,
\end{eqnarray}
where $\sigma_1=\sqrt{\langle|\nabla f|^2\rangle}$ is the rms of the
gradient of the field.  
For a given field whose statistical nature is not known a priori, one
can compute the MFs and compare with these analytic expressions to
test whether it is Gaussian or has non-Gaussian deviations. 

\subsection{Estimating MFs}

In Sects.~\ref{sec:3.1} and \ref{sec:3.2}, we have outlined our method
for constructing brightness temperature threshold masks using
$408$~MHz map and corresponding $1000$ Gaussian sky realisations for
the binned bispectrum estimation.  
To evaluate all the three MFs as a function of angular scales, we
apply a low-pass filtering to both actual and simulated data sets. 

\begin{equation}
  f(\ell) = \begin{cases}
    0 &  \text{for} \  \ell < \ell_{\rm c} - \Delta \ell \\
    \cos^{2}\left(\frac{\pi}{2} \times \frac{(\ell_{\rm c} - \ell)}{\Delta \ell}\right) & \text{for} \ \ell_{\rm c} - \Delta \ell\leq \ell < \ell_{\rm c}  \\
    1 & \text{for} \  \ell \ge \ell_{\rm c}. \\
  \end{cases} \label{eqn:v3}
\end{equation}
We choose three different values of low $\ell$ cutoff corresponding to
$l_{\rm c}= 50, 70, \text{and} \, 90$. 
Our filtering scheme takes care of artefacts arises in the filtered
$408$~MHz map due to sharp cut-off around the edge of the bin.
We checked that the final results is independent of the filtering
scheme that we employed in our analysis.
As in Sect.~\ref{sec:3.2} to restrain leakage from bright galactic
plane in binned filtered maps, we first apply the brightness
temperature threshold of $80$~K to $408$~MHz map.  

To compute MFs for an arbitrary field, we employ the method outlined
in \cite{Schmalzing1998}.
This method involves expressing Eq.~\eqref{eqn:v1} and~\eqref{eqn:v2}
regarding first and second derivatives of the field, converting the
line integrals to area integrals by introducing delta function for the
field at the threshold values, and then, integrating over the sphere. 

\begin{figure*}
  \includegraphics[width=14.5cm]{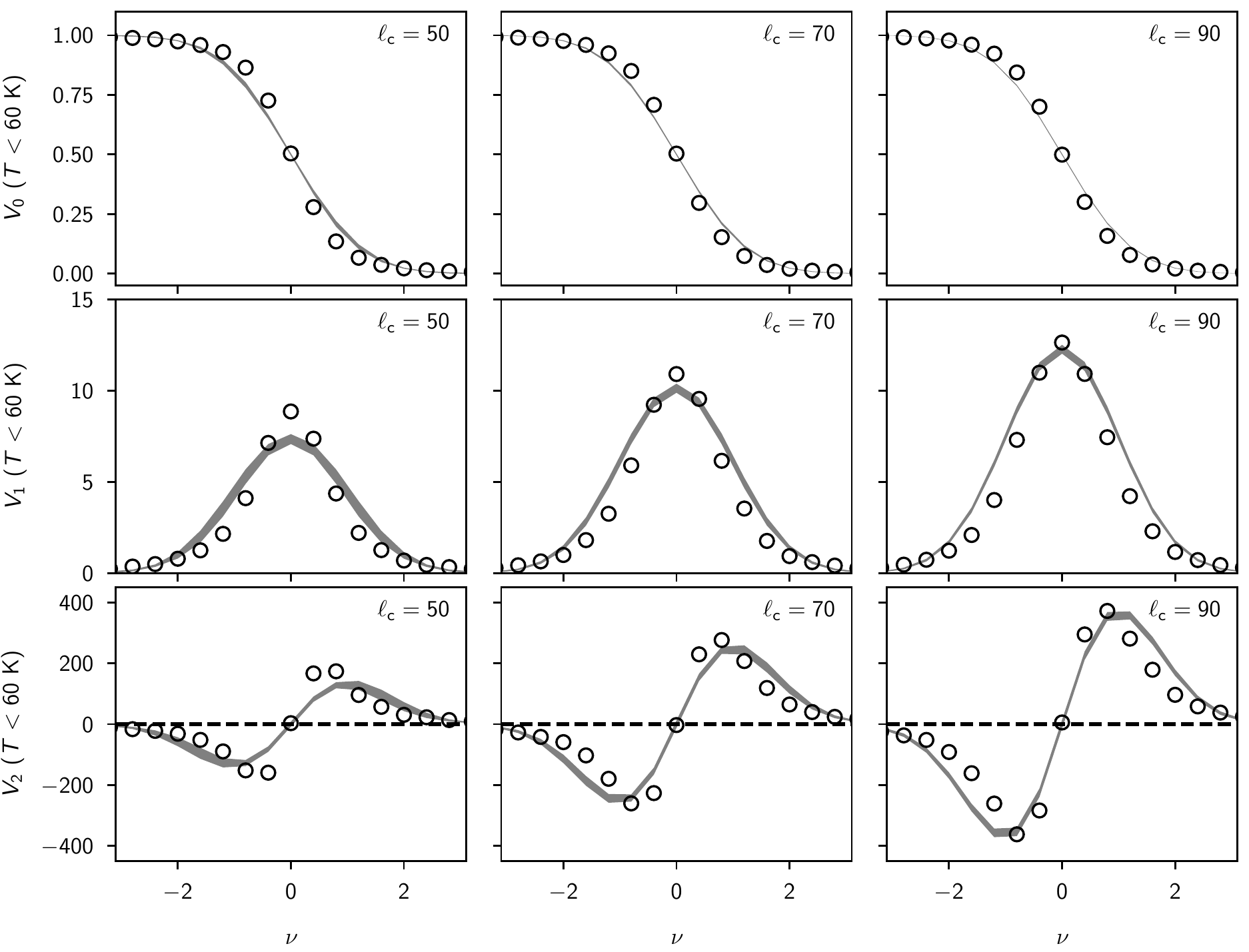}
  \caption {Black points indicate the variation of three MF for different 
    cut-off ($\ell_c$) in 408 MHz map with the sky brightness
    threshold, $T < 60$\,K.  
    Gray shaded region indicates 3-$\sigma$ deviation from the ensemble mean of 
    1000 Gaussian realisations. }
  \label{9}
\end{figure*}

\begin{figure*}
  \includegraphics[width=14.5cm]{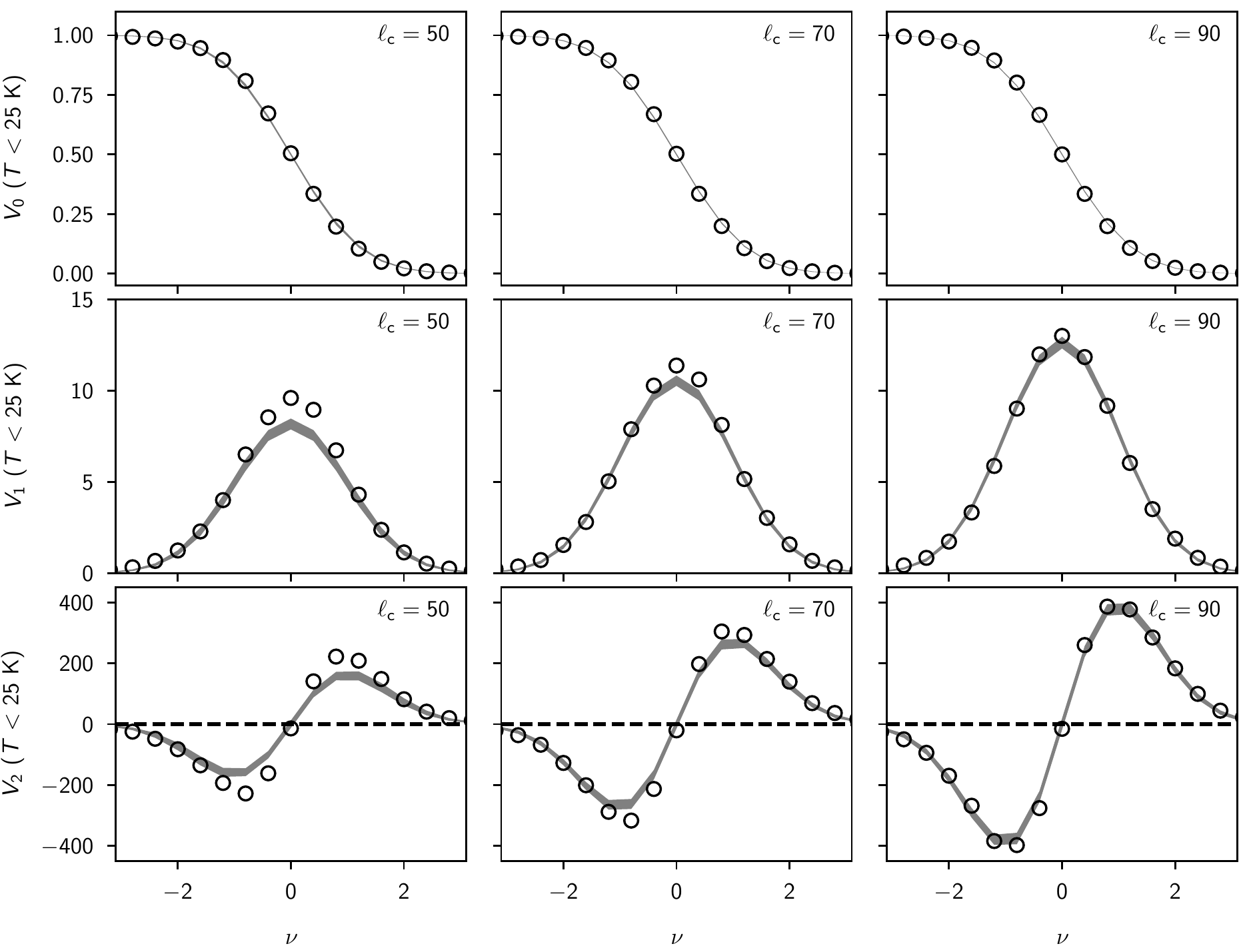}
  \caption {Same as Fig.~\ref{9}, but for 408~MHz brightness threshold
    $T < 25$\,K.  
    Both the data and Gaussian simulations, agree with each other well
    within 3-$\sigma$ for angular scales $\le 3\deg$.} 
  \label{8}
\end{figure*}

We use the same $1000$ random Gaussian realisations for each brightness
temperature threshold that we have generated for the binned bispectrum
analysis.  
We compute ensemble mean and $1\sigma$ standard deviation of three MF
for each of the four case.
For two extreme sky brightness thresholds $T<60$~K and $T<25$~K, we
plot the three MFs for the real data, and the ensemble mean of $1000$
Gaussian realisations as shown in Figs~\ref{9} and \ref{8}.
For $60$~K map, the real data and Gaussian simulations differ from
each other significantly for all three MFs.
Since MFs are sensitive to the entire hierarchy of higher order
correlations unlike bispectrum, therefore, we aspect higher deviation
($3\sigma$ grey shaded regions) between actual data and Gaussian
simulation.
As in case of bispectrum here also $25$~K maps, for multipole cut-off
$\ell_{c} \geq 60$ ($\theta < 3 \deg$), the real data is relatively
consistent with Gaussian simulations. 

\section{Summary and Discussion}

In this article, we discussed three-point statistics of Galactic
synchrotron emission at $408$~MHz.   
We used the re-processed version of all-sky $408$~MHz map and divide
the sky into four regions based on the sky brightness.
For each brightness temperature threshold using \polspice\ routine, we
obtained  $TT$ power spectra which are corrected for the sky mask,
beam smoothing and pixel window function and created $1000$ Gaussian
sky realisations.
We applied binned bispectrum estimator to both the data and Gaussian
realisations and obtained bispectrum value in bins of multipole. 
We have also applied MF estimator to compute three MF values for
different multipole cut-off $\ell_c$. Our conclusions are as follows: 
\begin{itemize}
\item
  Although the all-sky map of $408$~MHz sky is highly non-Gaussian, we 
  find that the regions with lower brightness temperature have a
  smaller bispectrum.
  For example, bispectrum in $408$~MHz map with a threshold at $25$~K
  is small enough to be indistinguishable from the ensemble of
  Gaussian simulations.
  For higher sky brightness temperature thresholds ($T > 40$~K), the
  data shows clear difference from the Gaussian sky realisations.  
\item
  The three MFs for the actual maps deviate significantly 
  from Gaussian realisations for higher brightness temperature thresholds.
  As we go to lower brightness threshold and higher value of 
  multipole cut-off ($\ell_c \geq 60$), the three MF of the real data 
  agree well the Gaussian simulations within the $3\sigma$.
\item
  However, the skewness and kurtosis of $a_{\ell m}$ distribution of
  $408$~MHz map agree well within $2\sigma$ with respect to Gaussian
  realisations for all sky brightness temperature thresholds. 
  Since skewness of $a_{\ell m}$ distribution corresponds to a
  equilateral configuration \citep{Pietrobon}, therefore it does not
  capture non-Gaussian features arising from other triangular
  configurations to which binned bispectrum estimator is sensitive. 
\end{itemize}
Our results indicate scales of $3\deg$ (or $\ell \geq 60$), and in
regions away from the Galactic plane, the statistics of the $408$~MHz
map is consistent with the Gaussian sky realisations. 
Our finding is consistent with the earlier results from \cite{Dav15a},
who reported that the Gaussian approximation is valid for scales
$3\pdeg7$ away from the Galactic plane.
Unlike binned bispectrum analysis, the MFs results are sensitive to
higher order correlations.
However, for low sky brightness threshold $T<25$~K MFs of the real
data agrees well within $3\sigma$ with the Gaussian simulations.
This indicates that for angular scales $\ell_{c} \geq 60$, the actual
data is consistent with the Gaussian simulations.
Our study validates the use of Gaussian approximation in Galactic
synchrotron simulations, as long as we restrict ourselves to cooler
parts of the sky and smaller angular scales.

In near future, we expect to have more of high-resolution full or
partial sky coverage low-frequency foreground maps from existing or
upcoming next-generation radio telescope facilities.
Studying higher order correlation on such maps will help us understand
and improve our foreground knowledge for future \eor\ experiments. 

%%%%%%%%%%%%%%%%%%%%%%%%%%%%%%%%%%%%%%%%%%%%%%%%%%%%%%%%%%%%%%%%%%%%%%%%%%%%%%%

\section*{Acknowledgement}

We acknowledge the use of HPC facility at IISER Mohali.
We thank Tarun Souradeep for helpful discussion on bispectrum theory. 
SR thanks Benjamin Racine and Biuse Casaponsa for helping out with
binned bispectrum estimator and code development and testing. 
The authors acknowledge NISER Bhubaneswar for funding academic visits
regarding the project and the use of their HPC facility. 
This research has made use of NASA Astrophysics Data System
Bibliographic Services.
Some of the results in this paper have been derived using the
\healpix\ \citep{Healpix} package. 

%%%%%%%%%%%%%%%%%%%%%%%%%%%%%%%%%%%%%%%%%%%%%%%%%%%%%%%%%%%%%%%%%%%%%%%%%%%%%%%%
%
\appendix

\section{Binned statistics} \label{appendix:A}

\begin{figure*}
\includegraphics[width=18.0cm,angle=0.0]{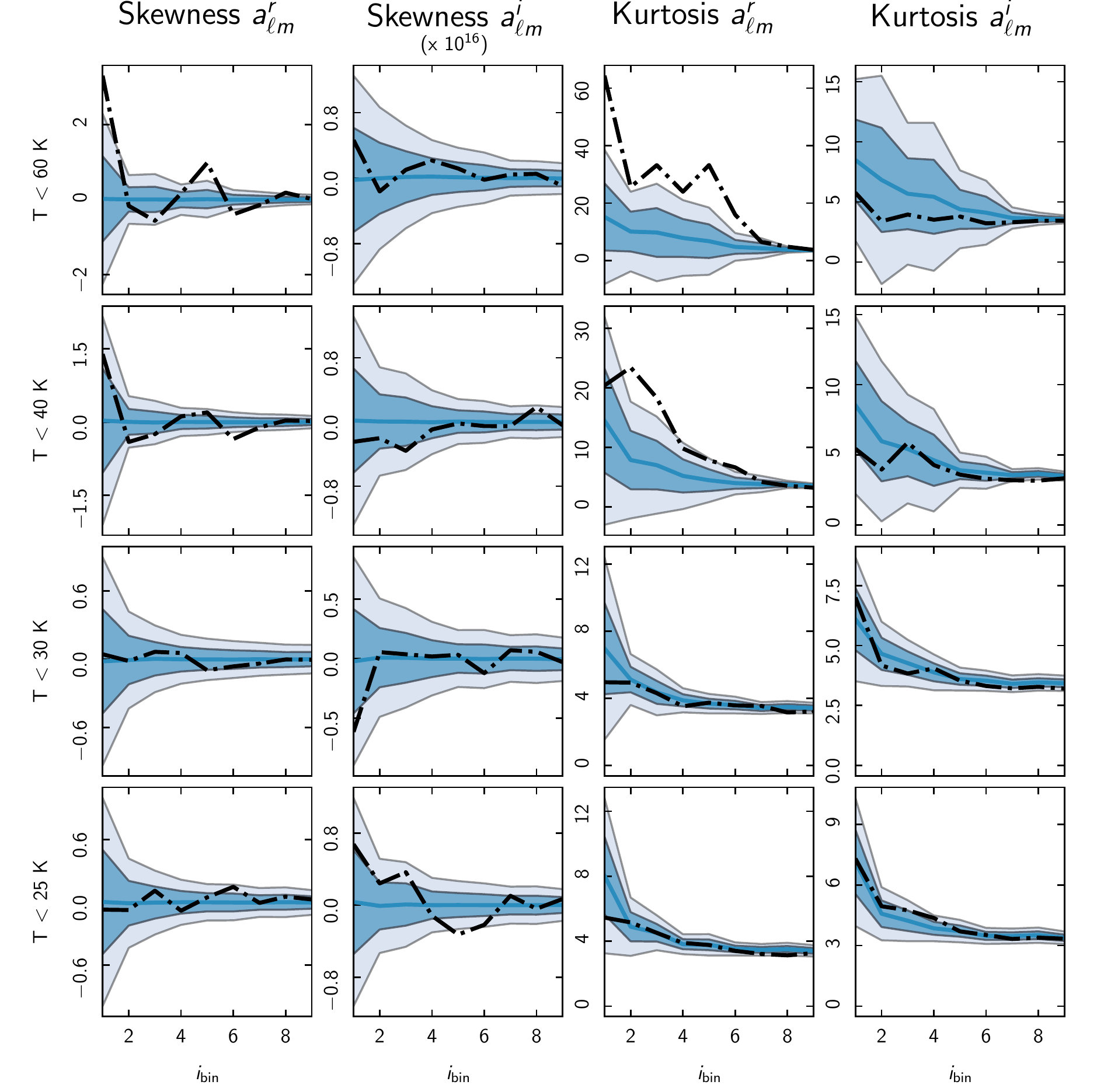}
\caption { $a_{\ell m}$ skewness and kurtosis versus $\rm i$ bins:
  light-blue line is mean value of $a_{\ell m}$ over 1000 Gaussian
  realisation in $\rm i$ bin and shaded regions are $1\sigma$ and
  $2\sigma$ confidence regions, black dashed line corresponds to
  actual map $a_{\ell m}$ value in $\rm i_{\rm bin}$.} 
\label{10}
\end{figure*}

In addition to estimating bispectrum, we also studied higher order
statistics such as skewness ($s$), and kurtosis ($\kappa$) of real
($a_{\ell m}^r$) and imaginary ($a_{\ell m}^i$) part of $a_{\ell m}$
distribution corresponds to different $\rm i$ bins.  
We then look how much they differ from each other in case of the
actual map, and corresponding Gaussian realisations (see
Fig.~\ref{10}).
Both skewness and kurtosis of $a_{\ell m}$ distribution in the real
data are in agreement with corresponding Gaussian realisations.
The skewness of $a_{\ell m}$ is sensitive to the equilateral
bispectrum configuration $\ell_1 \simeq \ell_2 \simeq \ell_3$
\citep{Pietrobon}. 

%%%%%%%%%%%%%%%%%%%%%%%%%%%%%%%%%%%%%%%%%%%%%%%%%%%%%

% Don't change these lines
%\bsp  % typesetting comment
%\label{lastpage}
\end{document}